\documentstyle[pre,aps,floats,epsf,twocolumn]{revtex}
\begin{document} 
\newcommand{\be}{\begin{equation}}
\newcommand{\ee}{\end{equation}}
\newcommand{\TS}[1]{{\tt <T>}{\bf #1}{\tt </T>}}
\newcommand{\AS}[1]{{\tt <A>}{\bf #1}{\tt </A>}}
\newcommand{\TAKENS}{t^{\rm ML}}
\newcommand{\BDS}{t^{\rm BDS}}
\newcommand{\PE}{t^{\rm PE}}
\newcommand{\CCC}{t^{\rm C3}}
\newcommand{\SKEW}{t^{\rm REV}}

\title{On the discrimination power of measures for nonlinearity in a time
   series}
\date{Phys.\ Rev.\ E {\bf 55}, 5443 (1997)}
\author{Thomas Schreiber  and Andreas Schmitz\\
      Physics Department, University of Wuppertal,\\
      D--42097 Wuppertal, Germany}
\maketitle
\begin{abstract} 
The performance of a number of different measures of nonlinearity in a time
series is compared numerically. Their power to distinguish noisy
chaotic data from linear stochastic surrogates is determined by Monte Carlo
simulation for a number of typical data problems. The main result is that the
ratings of the different measures vary from example to example. It seems
therefore preferable to use an algorithm with good overall performance, that
is, higher order autocorrelations or nonlinear prediction errors.
\end{abstract}

\section{Introduction}
The theory of nonlinear, deterministic dynamical systems provides powerful
theoretical tools to characterize geometrical and dynamical properties of the
attractors of such systems. Alongside the theoretical understanding of these
systems, many of the typical phenomena have been realized in laboratory
experiments. Many attempts have been made to detect behavior characteristic for
deterministic systems also in field data, that is, time series recordings of
real world phenomena. Not surprisingly, the coarse nature of these time series
(finite number of points with finite resolution) makes it difficult to obtain
unambiguous results. As a particular example, it has been pointed
out~\cite{oneoverf} that linear stochastic processes with long range
autocorrelations can lead to spuriously small estimates of the attractor
dimension. (See also the discussion in Ref.~\cite{nico}.) The method of
surrogate data~\cite{surro} provides a rigorous statistical test for the null
hypothesis that the data has been generated by a linear stochastic process.  If
this null hypothesis cannot be rejected, the results of a nonlinear analysis
have to be regarded as spurious. In such a test, the value of some measure of
nonlinearity is compared for the data and a number of randomized samples, the
surrogates. The nonlinearity measure should be sensitive to the kind of
nonlinearity suspected in the data and it should be possible to estimate its
value with low variance. In this paper we will numerically compare the
performance of a selection of measures which have been proposed in the
literature.

Apart from the mere detection of nonlinearity, nonlinear observables can be
used to discriminate between distinct states of a system on the base of time
series data. Most notably, claims have been made that measures derived from
chaos theory are able to distinguish healthy patients from those with
pathological biological rhythms, for example cardiac
arrhythmiae~\cite{moltofill,kurths,skinner}.  The results presented in this
paper are also of relevance for the question of the preferable discriminating
statistic in such a context.  The most striking observation is that although
the simplest observables, notably simple prediction errors, show good overall
performance, results differ immensely from application to application, which
may explain the partially contradicting claims in the literature. If enough
data is available to be split into a training set and a test set, and if a
model for a reasonable alternative hypothesis can be constructed, then
optimization of the test on typical data may be worthwhile.

\section{Testing for nonlinearity with surrogate data}
Currently, the most general null hypothesis we know how to test against is that
the data was generated by a stationary Gaussian linear stochastic process,
maybe measured through an instantaneous measurement function~\cite{we}.
Deviations from this null hypothesis can be detected by computing some
nonlinear observable on the data. Since the probability distributions of such
observables are generally not known analytically, they must be estimated by
Monte Carlo resampling of the data. For this purpose one generates random data
sets (surrogates) which conserve those properties of the data which are
irrelevant for a given choice of the null hypothesis. For the hypothesis of a
Gaussian linear stochastic process, the data and the surrogates must have the
same autocorrelation function or, equivalently, the same power spectrum. For a
nonlinearity test allowing for simple rescalings, also the single time
probability distribution must be conserved.  A (nonlinear) observable
$t=t(\{x_n\})$ is estimated on the original data $\{x^0_n\}$ and all of the $B$
surrogates $\{x^k_n\}, k=1,\ldots,B$. The distribution of $t$ can be estimated
from the values $t_k=t(\{x^k_n\})$. One can then test at a given level of
significance for the assumption that $t_0=t(\{x^0_n\})$ was drawn from the same
distribution. If this assumption is rejected, the original data $\{x^0_n\}$ is
taken to be different from the linear surrogates and is thus considered to be
nonlinear at this level of significance.

The use of surrogate data has been promoted in the context of chaotic time
series in Ref.~\cite{surro}.  Although the technique has made distinguishing
chaos from noise much safer, some caveats remain. These will not be discussed
in this paper, Refs.~\cite{sfi,tp,fields} provide noteworthy material.  We will
throughout use examples where the known problems (nonstationarity, long
coherence times) are of no concern.

There are two important parameters which characterize the performance of a
statistical test. One is its {\em size} $\alpha$, which is the probability that
the null hypothesis is rejected although it is in fact true. Specifying a {\em
level of significance} $1-p$ of the test amounts to the statement that its size
does not exceed $p$. It is customary to specify $p$ a priori and design the
test accordingly. The important question if the surrogate data test has indeed
the specified size has been previously addressed, see
Refs.~\cite{we,sfi,fields}. If the actual probability of a false rejection is
larger than $p$, the test yields incorrect results. The above references give
examples where this situation can occur with surrogate data tests.  While
excessive size renders the test useless, an actual size which is smaller than
$p$ is formally admissible. However, it can result in a dramatic decrease in
discrimination power. In such cases (for example if a fitted linear model is
run to generate surrogates), it is therefore advisable to calibrate the test by
using ``surrogate surrogate data''~\cite{fields}. Since the size of the test
may depend on the particular realization of the null hypothesis, this
calibration is usually quite cumbersome.  We verified the correct test size for
all the numerical examples in this paper by performing a series of tests on
surrogate data fulfilling the null hypothesis.

While the size predominantly assesses the quality of the surrogate data sets,
we want to evaluate in this paper the abilities of different {\em observables}
$t$ to detect nonlinearity. This property is quantified by the {\em power}
$\beta$ of the test. It is defined as the probability to correctly reject the
null hypothesis when it is indeed false.  The power of a statistical test can
be determined empirically by repeating the test many times on different
realizations of the data. Since we cannot make strong assumptions about the
distributions of the observables, there is no alternative to this
computationally expensive approach. In order to limit the computational effort,
however, we performed tests at a rather low level of significance, for which
only few surrogate data sets are necessary.

\section{Measures of nonlinearity}
We evaluated a number of different nonlinear observables. Most of them are
at least inspired by the theory of nonlinear dynamical systems and rely on a
time delay embedding of the scalar time series. Embedding vectors in $m$
dimensions are formed as usual: $\vec x_n=(x_{n-(m-1)\tau},\ldots,x_n)$, where
$\tau$ is the delay time. Since the Grassberger--Procaccia correlation
dimension $D_2$~\cite{GP} seems to be among the most popular measures we
considered several variants of this algorithm.  The correlation sum
$C(\epsilon)$ at a scale $\epsilon$ is given by
\be 
   C(\epsilon)={\rm const.}\times 
      \sum_{|i-j|>t_{\rm min}} \Theta(\|\vec{x}_i-\vec{x}_j\|-\epsilon)
\,.\ee
Dynamically correlated pairs are discarded as usual and $\rm const.$ refers to
the normalization. Since none of the examples in this study would allow for the
identification of a true scaling region, we will choose the length scales for
good discrimination power. Of course, this will make an interpretation as a
fractal dimension or complexity measure impossible.  In particular we
implemented two ways of turning $C(\epsilon)$ into a single number.
\begin{enumerate}
\item \label{t:takens} 
   A maximum likelihood (ML) estimator of the Grassberger--Procaccia 
   correlation dimension $D_2$ is given by:
   \be \TAKENS (m,\tau,\epsilon)=
     {C_m(\epsilon) \over 
     \int_0^{\epsilon}{C_m(\epsilon')\over\epsilon'}d\epsilon'}
   \,.\ee
   This expression is taken from Ref.~\cite{theiler88}. Maximum
   likelihood estimation of the correlation dimension goes back to
   Ref.~\cite{takens}. Therefore such quantities are generally referred to as
   {\em Takens' estimator}. 
\item \label{t:bds}
   Brock et al. (BDS)~\cite{brockpaper} have shown that for a sequence of
   independent random numbers, $C_m(\epsilon)=C_1(\epsilon)^m$ holds, where $m$
   is the embedding dimension.  In the same paper, also a formal test for this
   property is introduced. Instead of the original BDS statistic which has been
   introduced in order to be able to give the asymptotic form of the
   probability distribution, we use the simpler expression \be \BDS
   (m,\tau,\epsilon)=C_m(\epsilon)/C_1(\epsilon)^m \,. \ee
\end{enumerate}
Other choices we have tried are values of $C(\epsilon)$ at fixed length scales,
which gave consistently less power, and dimension estimators based on pointwise
dimensions. In the latter case, the scaling exponent of neighbor distances is
determined for each point separately. The actual observable is then the mean or
the median of these values~\cite{skinner,GMK}. Since we did not find any
interesting deviations from the power of the maximum likelihood estimator
$\TAKENS$ we did not include detailed results in this paper. 
   
Many quantities which have been proposed in the literature for nonlinearity
testing in some way or the other quantify the nonlinear predictability of the
signal. Examples include the statistic proposed by Kaplan and Glass~\cite{KG}
and to some extent also the false nearest neighbors techniques~\cite{FNN}. We
use a particularly stable representative of the class of predictability 
measures:
\begin{enumerate}
\setcounter{enumi}{2}
\item \label{t:pe} 
   A nonlinear prediction error with respect to a locally constant predictor
   $F$ can be defined by
   \be 
      \PE (m,\tau,\epsilon)=\left(\sum [x_{n+1}-F(x_n)]^2\right)^{1/2} 
   \,.\ee
   The prediction over one time step is performed by averaging over the future
   values of all neighboring delay vectors closer than $\epsilon$ in $m$
   dimensions.   
\end{enumerate}
In Ref.~\cite{nature} a nonlinear Volterra--Wiener model is claimed to be
superior to other techniques when applied to short noisy signals. We have
compared the maximal feasible noise level for a detection of nonlinearity
quoted in Ref.~\cite{nature} to the performance of the locally constant
predictor above for the H\'enon, Ikeda, and Lorenz series. We found that $\PE$
gave either better (H\'enon, Ikeda) or comparable (Lorenz) performance and
therefore did not include the Volterra--Wiener model in this study.

Further, we used the following nonlinear observables:
\begin{enumerate}
\setcounter{enumi}{3}
\item \label{t:c3}
   Linear (two point) autocovariances can be generalized by introducing more
   than one lag. In the spectral domain, this generalization leads to the
   bispectrum and polyspectra~\cite{BI}. Our (somewhat arbitrary) choice of a
   higher order autocovariance (or cumulant) is
   \be 
      \CCC (\tau)= \langle x_n x_{n-\tau} x_{n-2\tau} \rangle 
   \,.\ee
\item \label{t:rev} 
   A simple quantity which is frequently used to detect deviations from
   time-reversibility is
   \be 
      \SKEW (\tau)=\langle (x_n-x_{n-\tau})^3\rangle
   \,.\ee
\end{enumerate}

We have explicitly indicated the adjustable parameters which can be chosen
using several different strategies. One possibility is to optimize the
adjustable parameters. This has to be done either on data which is not
subsequently used for the test, or it has to be done individually for each data
set and surrogate. The former requires the knowledge of the correct answer for
the ``training data'' which is rather uncommon. The latter is computationally
extremely expensive and care has to be taken in order to avoid overfitting of
the data. Note that for example minimizing prediction errors does not
necessarily optimize the discrimination power.

In the present work, we fix as many parameters as possible to reasonable {\em
ad hoc} values prior to the tests. Before each test, a brief survey was
performed as to which embedding dimensions and delay times lead to satisfactory
results for each quantity. We feel that this procedure comes closest to what
one can do in practice, where also a formal optimization of the discrimination
power is impossible. The length scale $\epsilon$ was either determined as a
fixed fraction (1/4) of the root mean squared (\ref{t:takens}., \ref{t:bds}.)
or the peak-to-peak amplitude (\ref{t:pe}.) of the data. 

\section{Implementation and results}
The surrogate data sets will be generated as described in Ref.~\cite{we}, which
is the appropriate method when the null hypothesis is that the data has been
generated by a Gaussian linear stochastic process, possibly measured through a
monotonic, instantaneous, time independent measurement function. In brief, the
method is based on an ordinary phase randomized surrogate series $S=\{s_n,
n=1,\ldots,N\}$ which has the same sample power spectrum as the time series
$X=\{x_n, n=1,\ldots,N\}$. Such a surrogate is obtained by taking the Fourier
transform of $X$, randomizing the phases, and inverting the transform. Now the
following two steps are iterated alternatingly: 
\begin{enumerate}
\item The surrogate series is brought to the sample distribution of $X$
   by rank--ordering:
   \begin{equation}
      s'_n=x_{{\rm index}({\rm rank}(s_n))}
   \,,\end{equation}
   Here, ${\rm rank}(s_n)=k$ and ${\rm index}(k)=n$ if $s_n$ is the $k$-th
   smallest value in $S$. After this step, $S'$ and $X$ have the same
   distribution of values, but the power spectrum may have changed.
\item The Fourier amplitudes of $S'=\{s'_n, n=1,\ldots,N\}$ are replaced by
   those of $X$. The resulting series $S''$ has the same sample power spectrum
   as $X$. This step may however alter the distribution of values.
\end{enumerate}
In Ref.~\cite{we} numerical evidence and heuristic arguments are given that
this scheme indeed converges to a sequence with the same distribution {\em and}
the same power spectrum as the data. While formal convergence can only be
expected for infinitely long sequences, the approximation is satisfactory for
finite data length. If the deviation from a Gaussian distribution or the linear
correlations in the time series are not too strong, the usual amplitude
adjusted phase randomized surrogates~\cite{surro} yield an accurate test as
well. Our results do not explicitly depend on the particular method of
generating constrained Monte Carlo realizations.

As mentioned before, we do not know the probability distributions of the
nonlinear observables used in this paper. In particular, Gaussianity cannot be
assumed. Therefore we have to employ a non-parametric, rank-based test, as it
has been suggested in Ref.~\cite{tp}. A test is called {\em one-sided} if the
null hypothesis is rejected only if the data deviates from the surrogates in a
specified direction. In this case and at a given size $\alpha$, we create
$B=1/\alpha-1$ surrogate data sets and compute the test statistic $t_0$ on the
original data set and its value $t_k, k=1,\ldots,B$ on each of the
surrogates. Since we have a total of $1/\alpha$ sets, the probability for each
of them to have the smallest value of $t$ by chance is just $\alpha$, as
desired. For two-sided tests, we generate $B=2/\alpha-1$ surrogates. The
probability for any of the $2/\alpha$ sets to have either the smallest or the
largest value of $t$ is then again $\alpha$.

For the nonlinearity measures inspired by the theory of deterministic dynamical
systems (\ref{t:takens}.--\ref{t:pe}. above), we expect nonlinearity in the
data to result in lower values. Thus it is natural to perform
one-sided tests. For the remaining two measures we perform two-sided tests.  In
order to limit the computational burden, all tests are carried out at the 90\%
level of significance, that is, with 9 (resp. 19 for two-sided tests)
surrogates. For practical applications, at least a 95\% confidence is usually
required. The power can be increased by performing tests based on more than the
minimal number of surrogate data sets.

For purely deterministic signals, we would almost invariably get a
discrimination power of $\beta=1$. Therefore we contaminate deterministic
sequences $\{x_n\}$ with noise $\{\eta_n\}$ which consists of a phase
randomized copy of the sequence.  Thus the noise is random but with the same
power spectrum as the data ({\em in-band noise}). The noisy data is given by:
\be 
   s_n=\sqrt{\frac{1}{1+a^2}}(x_n+a\eta_n)
\,.\ee
The way the noise is generated and added guarantees that the power is not
dominated by changes in the autocorrelations or the variance of the data.

\begin{figure}
\centerline{\epsffile{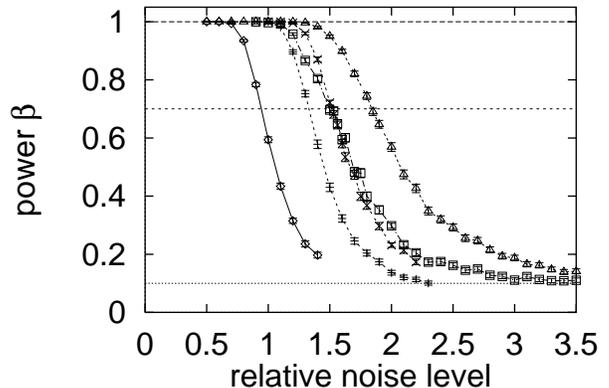}}
\caption[]{\label{fig:henon}
   Comparison of discrimination power for different nonlinearity measures and
   noise levels for H\'enon data with in-band noise. Curves from the left:
   correlation statistics $\TAKENS$ and $\BDS$, prediction error $\PE$
   (crosses), third order cumulant $\CCC$, time asymmetry $\SKEW$.  The size of
   the test was taken to be 0.1.}
\end{figure}

One sequence of tests is performed at different noise levels in order to
determine the maximal feasible noise level which allows for the detection of
nonlinearity with a power of $\beta=0.95$ resp. $\beta=0.7$. The practical
usefulness of tests with power less than $\beta=0.7$ seems questionable.  In
this sequence, 2000 individual tests with H\'enon time series of length 2048
were carried out for each point. The results are summarized in
Table~\ref{tab:henon} and Fig.~\ref{fig:henon}. For this discrete time system,
unit time delay seems most appropriate.

\begin{table}
\begin{center}
\begin{tabular}{llcc}
\hline\hline
statistic & parameters & 
         \multicolumn{2}{c}{feasible noise level $a_{\rm max}$} \\
          & & $\beta=0.95$ & $\beta=0.7$ \\
\hline
$\TAKENS$       & $m=2$                 & 0.7   & 0.9\\
$\BDS$          & $m=3$                 & 1.1   & 1.3\\
$\PE$           & $m=3$                 & 1.2   & 1.5\\
$\CCC$          & $\tau=1$   & 1.1   & 1.5\\
$\SKEW$         & $\tau=1$              & 1.4   & 1.8\\
\hline\hline
\end{tabular}
\end{center}
\caption{\label{tab:henon}
Maximal feasible noise level for the detection of nonlinearity with
$\beta=0.95$ resp. $\beta=0.7$. Results for the H\'enon map.}
\end{table}

Further, we evaluated the different quantities for a number of particular data
problems, time series from the Lorenz equations, an NMR laser experiment, and
an assembly of uncoupled tent maps. In Table~\ref{tab:lorenz} we show the
results for time series of the Lorenz system at standard parameter values. 2048
samples of the $x$-coordinate were recorded every 0.08 time units. Noise of
amplitude $a=1.3$ was added.  It was checked for each of the different
observables (but with fewer tests) that other choices of the lag time and the
embedding dimension did not lead to significantly better results.

\begin{table}
\begin{center}
\begin{tabular}{lll}
\hline\hline
statistic & parameters & power $\beta$\\
\hline
$\TAKENS$       & $m=3$                 & 0.25 $\pm$ 0.02 \\
$\BDS$          & $m=2$                 & 0.24 $\pm$ 0.02 \\
$\PE$           & $m=4$                 & 0.66 $\pm$ 0.02 \\
$\CCC$          & $\tau=3$   & 0.09 $\pm$ 0.01 \\
$\SKEW$         & $\tau=3$              & 0.10 $\pm$ 0.01 \\
\hline\hline
\end{tabular}
\end{center}
\caption{\label{tab:lorenz}
Fraction of successful rejections out of 1000 tests, noisy Lorenz data. The
errors are based on the assumption of a binomial distribution for independent
trials. No significant rejection is possible with $\CCC$ and $\SKEW$.}
\end{table}

A long experimental time series from an NMR laser experiment~\cite{NMR} was
split into 600 segments with 1000 points each. In-band noise of amplitude
$a=0.8$ was added. Results are shown in Table~\ref{tab:raser}.

\begin{table}
\begin{center}
\begin{tabular}{lll}
\hline\hline
statistic & parameters & power $\beta$\\
\hline
$\TAKENS$       & $m=3$                 & 0.61 $\pm$ 0.03 \\
$\BDS$          & $m=3$                 & 0.86 $\pm$ 0.02 \\
$\PE$           & $m=3$                 & 0.79 $\pm$ 0.02 \\
$\CCC$          & $\tau=3$   & 0.45 $\pm$ 0.03 \\
$\SKEW$         & $\tau=1$              & 0.35 $\pm$ 0.02 \\
\hline\hline
\end{tabular}
\end{center}
\caption{\label{tab:raser}
Fraction of successful rejections out of 600 tests, noisy NMR laser data.}
\end{table}

\begin{figure}
\centerline{\epsffile{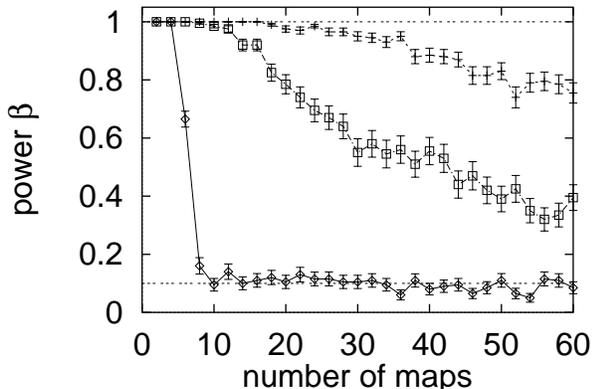}}
\caption[]{\label{fig:maps}
   Discrimination power for uncoupled tent maps. In this figure, results are
   shown for three selected nonlinearity measures, from above: time asymmetry
   $\SKEW$, prediction error $\PE, m=4$, and ML dimension estimator $\TAKENS,
   m=2$. The number of maps was varied in steps of two, each point was obtained
   with 200 tests. The size of the test was taken to be 0.1.}
\end{figure}

Finally, we consider an assembly of uncoupled tent maps. Each individual map is
given by $x_{n+1}=2x_n$ if $x<0.5$ and $x_{n+1}=2-2x_n$ if $x\ge 0.5$.  The
recorded variable is the sum of the variables of $N$ individual tent maps. No
noise is added. The discrimination power is measured as a function of $N$. In
Fig.~\ref{fig:maps} we show the results for the time asymmetry, prediction
error, and ML statistics. Table~\ref{tab:maps} shows the results for all the
nonlinearity measures at $N=16$. In this example, the time asymmetry statistic
is doing extremely well. The prediction error also gives reasonable
power while all other quantities basically fail, although different settings
for $m$ were considered.
\begin{table}
\begin{center}
\begin{tabular}{lll}
\hline\hline
statistic & parameters & power $\beta$\\
\hline
$\TAKENS$       & $m=2$                 & 0.11 $\pm$ 0.03 \\
$\BDS$          & $m=2$                 & 0.10 $\pm$ 0.03 \\
$\PE$           & $m=4$                 & 0.92 $\pm$ 0.03 \\
$\CCC$          & $\tau=1$   & 0.10 $\pm$ 0.03 \\
$\SKEW$         & $\tau=1$              & 1.00 $\pm$ 0.00 \\
\hline\hline
\end{tabular}
\end{center}
\caption{\label{tab:maps}
Fraction of successful rejections out of 100 tests, sum of  16 uncoupled
tent maps.}
\end{table}

\section{Conclusions}
The results presented in Tables~\ref{tab:henon}--\ref{tab:maps} and the figures
suggest that the root mean squared error of a simple nonlinear predictor gives
consistently good discrimination power. Other nonlinearity measures give even
better performance in some cases, but fail in others. In particular, the time
reversal asymmetry does very well most of the time but can also fail
completely.  Asymmetry under time reversal is a sufficient and powerful
indicator of nonlinearity, but not a necessary condition. Which algorithm is to
be preferred in a particular situation depends on the availability of an
independent check for the discrimination power.  In the typical situation that
only few precious data sets, or even just one recording (like in long-term
geophysical observations) is available, it seems advisable to use a robust,
general purpose statistic with few adjustable parameters, for example a simple
prediction error. If asymmetry under time reversal appears under visual
inspection of the data, a simple statistic like $\SKEW$ will probably give best
results.

The null hypothesis we have adopted in this work has been chosen since it is
the most general one that excludes nonlinear determinism and that can be tested
for properly. If we are in fact looking for deterministic structure in a
signal, then simple statistics like $\SKEW$ and $\CCC$ which are based on
higher order cumulants are not very attractive because they are quite sensitive
also to those deviations from the null hypothesis we are {\em not} looking
for. The formal test discussed in this paper answers the question if {\em any}
deviation from a (rescaled) Gaussian linear stochastic process can be
detected. Surrogate data tests have however been mostly used with the question
in mind if it is legitimate and useful to use methods from dynamical systems
theory. This amounts to specifying a particular class as an alternative
hypothesis. In such a case we should choose the discriminating statistic
accordingly, that is from the arsenal of dynamical systems methods.

Let us finally remark that a couple of tests for nonlinear properties of time
series have been proposed which use surrogate data in a different way or not at
all.  Rather than estimating the distribution of the observable $t$ from a
randomized sample, it is sometimes calculated on the base of some
assumptions. If the null hypothesis is that of a purely Gaussian linear random
process (without distortion), significance levels for higher order correlation
functions can be derived. Some authors, e.g. Ref.~\cite{skinner,nature},
observe that most observables $t$ are temporal averages over individual
quantities $t_n$ determined for each point in a time sequence. In order to
derive the distribution of $t$ from the knowledge of $\{t_n\}$ however, one has
to make certain assumptions. Ref.~\cite{skinner} assumes Gaussianity and
independence of the $t_n$ while Ref.~\cite{nature} only needs independence.  We
do not see what should justify the assumption of point-to-point independence of
$\{t_n\}$ for autocorrelated time series data, indeed we empirically find the
assumption to be wrong at least for prediction errors and pointwise
dimensions. The common positive correlation among the $t_n$ leads to an
underestimation of the variance of the average $t$ and thus to a dangerous
overestimation of the significance of the test.

\section*{Acknowledgments}
We thank James Theiler, Daniel Kaplan, Peter Grassberger, and Holger
Kantz for stimulating discussions.  This work was supported by the SFB 237 of
the Deutsche Forschungsgemeinschaft.

\end{document}